\documentclass[intlimits,twoside,a4paper]{article}
\usepackage[cp1251]{inputenc}

\usepackage{cmpj3}

\usepackage{bm}

\issue{2021}{24}{1}{13705}
\doinumber{10.5488/CMP.24.13705}
\title[Renormalized spectrum of quasiparticle in limited number of states, strongly interacting\ldots]%
{Renormalized spectrum of quasiparticle in limited number of states, strongly interacting with two-mode polarization phonons at $T=0$~K}
\author[M.V.~Tkach, Ju.O.~Seti, O.M.~Voitsekhivska, V.V.~Hutiv]{M.V.~Tkach \footnote{E-mail: m.tkach@chnu.edu.ua},
Ju.O.~Seti, O.M.~Voitsekhivska, V.V.~Hutiv}
\address{Yuriy Fedkovych Chernivtsi National University, 2 Kotsyubinsky Str., 58012 Chernivtsi, Ukraine}

\date{Received August 10, 2020, in final form September  17, 2020}
\begin{document}

\maketitle

\begin{abstract}
Within unitary transformed Hamiltonian of Fr\"{o}hlich type, using the Green's functions method, exact renormalized energy spectrum of quasiparticle strongly interacting with two-mode polarization phonons is obtained at $T=0$~K in a model of the system with limited number of its initial states. Exact analytical expressions for the average number of phonons in ground state and in all satellite states of the system are presented. Their dependences on a magnitude of interaction between quasiparticle and both phonon modes are analyzed.
\end{abstract}

\section{Introduction}
Just after the beginning of intensive development of the theory of quasiparticles (electrons, holes, excitons, impurities, etc.) interacting with phonons, when states with energies close to the energies of uncoupling quasiparticles were mainly studied, from physical considerations it became clear that different complexes of bound-to-phonons states are observed in the vicinity of threshold energies (order of quasiparticle energy plus one or more phonons).

At first, the theory of near-threshold phenomena was developed using the Hamiltonians of Fr\"{o}hlich type for the systems of quasiparticles interacting with phonons. The analytical calculations were performed within the methods of quantum field theory \cite{Abr12,Mah81,Lev74}. However, since the range of energies that did not exceed the radiation threshold of one phonon was usually observed, the investigations were provided in one-phonon approximation for the mass operator. The higher corrections to the complete vertex component were taken into account rarely. The theoretical results obtained during this period are presented in review  \cite{Lev74}. Here, the classification (over the average number of phonons   in the so-called phonon ``dress'' of quasiparticle) was proposed for the renormalized, due to interaction with phonons, three types of states of impurity centers: polaron ($N<1/2$), hybrid  ($N\sim1/2$), corresponding to the dielectric modes, known from experiments and bound-to-phonons states of quasiparticle  ($N\sim1$).

While the theory developed, it turned out that multi-phonon processes play an important role in the formation of renormalized spectrum of the system, when not only ground but also high-energy states were taken into account \cite{Mis05,Mis09,Dev09}. Within a universal method of Feynman diagram technique for the Fourier image of quasiparticles Green's function, it was established that sequential consideration of only a certain number of first components of complete mass operator (identical to perturbation theory), the same as infinite partial summing of all diagrams without crossing the phonon lines (due to the factorial decrease of their number compared to the rest diagrams with crossing lines in each order) did not provide a correct account of multi-phonon processes.

At the edge of 2000th, non-perturbative approaches and methods appeared, which correctly took into account multi-phonon processes in the theory of quasiparticle interacting with phonons: exact diagonalization (ED) for small systems, variational method (VM), the momentum average (MA) approximation and so on \cite{Bon99,Fil05,Bie16,Ber06,Cov07,Kor06,Ebr12,Mol16,Mar17}. Diagrammatic Monte Carlo (DMC) and bold diagrammatic Monte Carlo (BDMC) methods  \cite{Pro98, Mis00,Mar10,Fil12,Mis14} have played a particularly important role for investigation of high-excited states, revealing the reasons for different phenomena in electron-phonon systems. Both methods are essentially based at the computer algorithms for calculating the high-order diagrams of mass operator in Matsubara's Green's functions.

At the background of an intensive development of general theory for the systems of quasiparticles interacting with one-mode phonons, the multi-mode ones started to be studied long ago \cite{Agr68,Dav76}. However, the lack of an urgent need in explaining the key experimental phenomena and the emerging novel mathematical problems did not contribute to a sufficient development of such a theory.

Situation has essentially changed due to the recent rapid development of experimental and theoretical nano physics \cite{Gio09,Str01}. Strong spatial confinement of constituent elements of nano-heterostructures causes the appearance of different modes of phonons (confined, half-space, interface and propagating) interacting with different quasiparticles (electrons, holes, excitons, impurities, etc.). Even in the simplest nanostructure with one quantum well embedded into barrier-medium, for example HgS/CdS, the energies of two modes of polarization phonons are $\Omega_{\textmd{HgS}} = 27.8$~meV  and  $\Omega_{\textmd{CdS}} = 57.2$~meV, respectively. Although there is an urgent need, we still do not have a consistent and effective theory of quasiparticles interacting with multi-mode phonons.

It should be mentioned that in \cite{Dav76}, one of the approaches to the study of a renormalized spectrum of the two-state quasiparticle strongly interacting with polarization phonons was proposed. The retarded Green's function was exactly calculated in general form within the unitary transformed Hamiltonian of Fr\"{o}hlich type. However, the exact Fourier transformation and, hence, the renormalized discrete spectrum was obtained only for the system with one-phonon mode. An approximately calculated \cite{Dav76} spectrum of the system with an arbitrary number of phonon modes revealed to be decaying as at $T\neq0$~K, so at $T=0$~K. Such a result is evidently an artefact of the approximation because, from physical considerations, the interaction only with virtual phonons ($T=0$~K) cannot cause any decay.

A renormalized spectrum of localized quasiparticle interacting with one-mode dispersionless phonons at $T=0$~K was  obtained later, using the Feynman diagram technique \cite{Tka84,Tka16} and momentum average approximation \cite{Ber06}, for the model of the system without limiting conditions. It was the same as in \cite{Dav76}.

In this paper, within the exact calculation of Fourier image of retarded Green’s function, we obtained a renormalized spectrum of quasiparticle strongly interacting with dispersionless two-mode phonons at $T=0$~K in the model with limiting conditions. The average phonon numbers are analyzed for the main and satellite states of the system. The proposed method of calculation of Fourier image of retarded Green’s function opens up prospects to generalize it for the systems of localized quasiparticles interacting with an arbitrary number of phonon modes.

\section{Renormalized spectrum of the system and average numbers of pho\-nons in all states}
The system which consists of one-band quasiparticle (exciton, impurity etc.) strongly interacting with dispersionless two-mode phonons is studied at $T=0$~K. Its Hamiltonian is written in Fr\"{o}hlich form \cite{Dav76}
\begin{equation} \label{eq1}
\hat{H} = \sum_{\vec{k}} E_{\vec{k}} \hat{A}^+_{\vec{k}} \hat{A}_{\vec{k}} + \sum_{\lambda=1}^2 \sum_{\vec{q}} \Omega_{\lambda} \hat{B}^+_{\lambda\vec{q}} \hat{B}_{\lambda\vec{q}} + \sum_{\lambda=1}^2 \sum_{\vec{k}, \vec{q}} \varphi_{\lambda}(\vec{q}) \hat{A}^+_{\vec{k}} \hat{A}_{\vec{k}} \left(\hat{B}_{\lambda\vec{q}} + \hat{B}^+_{\lambda-\vec{q}} \right). \end{equation}
Here, $E_{\vec{k}}$ is an energy of uncoupled quasiparticle, $\Omega_{\lambda}$ is an energy of  $\lambda$-th phonon mode, $\varphi_{\lambda}(\vec{q})$ is a binding function. Quasiparticles  $ (\hat{A}^+_{\vec{k}},\hat{A}_{\vec{k}})$ and phonons  $(\hat{B}^+_{\lambda\vec{q}},\hat{B}_{\lambda\vec{q}})$ operators of second quantization satisfy Bose commutative relationships. Like in \cite{Dav76}, we are using the model for the system with the condition
\begin{equation} \label{eq2}
\hat{n}^{2}=\hat{n}=\sum_{\vec{k}}\hat{A}^+_{\vec{k}}\hat{A}_{\vec{k}} \,.
\end{equation}
It means that the eigenvalues of both these operators ($\hat{n}$ and $\hat{n}^{2}$) are either 0 or 1 and are interpreted as the condition of presence (1) or absence (0) of ``pure'' quasiparticle state.

To obtain a spectrum of such a system, the Hamiltonian (\ref{eq1}) is diagonalized like in \cite{Dav76} within transition from $\hat{A}_{\vec{k}}, \hat{B}_{\lambda\vec{q}}$ operators to a new  $\hat{a}_{\vec{k}}, \hat{b}_{\vec{q}\lambda}$ using a unitary operator
\begin{equation} \label{eq3}
S=\exp \displaystyle \{\hat{\sigma} \sum_{\vec{k}} \hat{a}_{\vec{k}} \hat{a}^+_{\vec{k}} \},
\end{equation}
where
\begin{equation} \label{eq4}
\hat{\sigma} = \sum_{\lambda, \vec{q}} \Omega^{-1}_\lambda \left[\varphi^{*}_\lambda(q) \hat{b}^+_{\lambda\vec{q}} - \varphi_\lambda(q) \hat{b}_{\lambda\vec{q}} \right].
\end{equation}
As a result, Hamiltonian  (\ref{eq1}) gets a diagonal form in new operators
\begin{equation} \label{eq5}
\hat{H}=\sum_{\vec{k}} \varepsilon_{\vec{k}}  \hat{a}^+_{\vec{k}}\hat{a}_{\vec{k}} + \sum_{\lambda, \vec{q}} \hat{b}^+_{\lambda\vec{q}} \hat{b}_{\lambda\vec{q}}\,,
\end{equation}
where
\begin{equation} \label{eq6}
\varepsilon_{\vec{k}} = E_{\vec{k}} - \sum_{\lambda, \vec{q}} \Omega^{-1}_{\lambda}\left| \varphi_{\lambda}(\vec{q}) \right|^{2}
\end{equation}
is an energy of new elementary excitations created by operator $\hat{a}^+_{\vec{k}}$.

Introducing (at $T=0$~K) a two-time retarded Green's function
\begin{equation} \label{eq7}
G(\vec{k}, t) = - \ri \theta(t)\left <0 \left | \left [\hat{A}_{\vec{k}}(t), \hat{A}^+_{\vec{k}}(0) \right] \right |0 \right>,
\end{equation}
taking into account (\ref{eq5}) and the relationship between old and new operators
\begin{equation} \label{eq8}
A_{\vec{k}} = S \hat{a}_{\vec{k}} S^{+} = \hat{a}_{\vec{k}} \re^{-\hat{\sigma}}
\end{equation}
and using the Weyl operator equality \cite{Dav76}, we obtain an exact expression
\begin{equation} \label{eq9}
G(\vec{k} ,t) = - \ri \theta(t) \left <0 \left | \re^{\hat{\sigma}^{+}(t)} \re^{\hat{\sigma}(0)} \right |0 \right> \exp\left (-\frac{\ri \varepsilon_{\vec{k}} t}{\hbar} \right) = - \ri \theta(t) \exp\left(-\frac{\ri \varepsilon_{\vec{k}} t}{\hbar} + g(t) \right).
\end{equation}
Here,
\begin{equation} \label{eq10}
g(t) = \sum_{\lambda=1}^2 \alpha_{\lambda} \left \{\exp \left (-\frac{\ri \Omega_{\lambda} t}{\hbar} \right) - 1\right\},
\end{equation}
where
\begin{equation} \label{eq11}
\alpha_{\lambda} = \Omega^{-2}_{\lambda} \sum_{\vec{q}} |\varphi_{\lambda}(\vec{q})|^{2}
\end{equation}
is a dimensionless parameter, which characterizes the binding energy of quasiparticle with  $\lambda$-th phonon mode.

Now, the Fourier image of Green's function (\ref{eq9}) is written in the form
\begin{equation} \label{eq12}
G(\vec{k}, \omega + \ri \eta) = - \frac{\ri}{\hbar} \int^{\infty}_{0} \exp \left \{\ri \left ( \omega - \frac{\varepsilon_{k}}{\hbar} + \ri \eta \right ) t + \sum_{\lambda=1}^{2} \alpha_{\lambda} \left [ \exp \left (- \frac{\ri \Omega_{\lambda} t}{\hbar} \right) - 1 \right ] \right \} \rd t.
\end{equation}

Expanding $\exp[\sum\limits_{\lambda=1}^{2} \alpha_{\lambda} \exp(-\ri \Omega_{\lambda} t / \hbar)]$ into a series and using Newton's binomial \cite{Gra84}, integral (\ref{eq12}) is calculated exactly. As a result, at $\hbar=1$, we obtain
\begin{equation} \label{eq13}
G(\vec{k}, \omega + \ri \eta) = \re^{-\sum\limits_{\lambda=1}^{2} \alpha_{\lambda}} \sum_{p=0}^{\infty} \sum_{l=0}^{p} \frac{\alpha^{p-l}_{1} \alpha^{l}_{2}} {(p-l)! \, l! \left[\omega - E_{k} + \sum\limits_{\lambda=1}^{2} \alpha_{\lambda} \Omega_{\lambda} - (p-l) \Omega_{1} - l \Omega_{2} + \ri \eta\right]}.
\end{equation}
Changing the order of summing in (\ref{eq13}), we again obtain  exact, but more convenient  for physical analysis, representation for Fourier image of Green's function
\begin{equation} \label{eq14}
\begin{split}
G(\vec{k}, \omega + \ri \eta)& = \re^{-\sum\limits_{\lambda=1}^{2} \alpha_{\lambda}} \left\{\frac{1}{\omega - E_{k} + \sum\limits_{\lambda=1}^{2} \alpha_{\lambda} \Omega_{\lambda}+\ri \eta} + \sum_{\lambda=1}^{2} \sum_{l_{\lambda}=1}^{\infty} \frac{\alpha_{\lambda}^{l_{\lambda}}} {l_{\lambda}! \left [\omega - E_{k} + \sum\limits_{\lambda_{1}=1}^{2} \alpha_{\lambda_{1}} \Omega_{\lambda_{1}} - l_{\lambda} \Omega_{\lambda} + \ri \eta\right]}\right.\\
&\qquad \qquad \left. + \sum_{l_{1}, l_{2}=1}^{\infty} \frac{\alpha_{1}^{l_{1}} \alpha_{2}^{l_{2}}}{l_{1}! \, l_{2}! \left [\omega - E_{k} + \sum\limits_{\lambda_{1}=1}^{2} \alpha_{\lambda_{1}} \Omega_{\lambda_{1}} - l_{1} \Omega_{1} - l_{2} \Omega_{2} + \ri \eta \right]} \right\}.
\end{split}
\end{equation}

According to the general theory \cite{Abr12,Agr68,Dav76}, at $T=0$~K the retarded and casual Green’s functions are the same, since the poles of $G(\vec{k}, \omega + \ri \eta)$ exactly determine the energy spectrum of the system. Thus, energetic denominators of three components in formula (\ref{eq14}) show that the renormalized energy spectrum can be expressed within one exact analytical expression
\begin{equation} \label{eq15}
E_{l_{1}, l_{2}}(\vec{k}) = E_{\vec{k}} - \sum_{\lambda=1}^{2} \alpha_{\lambda} \Omega_{\lambda} + l_{1} \Omega_{1} + l_{2} \Omega_{2}, \qquad (l_{1}, l_{2} = 0, 1, 2, \dots, \infty).
\end{equation}

From this formula it is clear that as far as there is no decay  $(\eta\rightarrow0)$, then, the spectrum of the system is real. Its main renormalized band (at $l_{1}=l_{2}=0$)
\begin{equation} \label{eq16}
E_{0, 0}(\vec{k}) = E_{\vec{k}} - \sum_{\lambda=1}^{2} \alpha_{\lambda} \Omega_{\lambda}
\end{equation}
is shifted into low-energy range, with respect to the main band $(E_{\vec{k}})$ of uncoupled quasiparticle, at the magnitude $\sum\limits_{\lambda=1}^{2}\alpha_{\lambda}\Omega_{\lambda}$. Besides, there is an  infinite number of discrete groups of satellite levels, which correspond to the bound states of quasiparticle with all possible combinations of a different number of phonons of both modes. They can be classified: a) two infinite unmixed groups of equidistant levels at $E_{l_{1}} ~ (l_{2}=0;\,l_{1}=1,2,\dots)$ and $E_{l_{2}} ~ (l_{1}=0; \,l_{2}=1,2,\dots)$
\begin{equation} \label{eq17}
E_{l_{\lambda}}(\vec{k} = 0) = E_{0} - \sum_{\lambda_{1}=1}^{2} \alpha_{\lambda_{1}} \Omega_{\lambda_{1}} + l_{\lambda} \Omega_{\lambda}, \qquad (l_{\lambda} = 1, 2, \dots, \infty; \,\, \lambda = 1, 2),
\end{equation}
which correspond to all possible bound states of a quasiparticle split by the energies as each phonon mode  b)~infinite number of mixed groups of satellite levels
\begin{equation} \label{eq18}
E_{l_{1}, l_{2} \neq 0}(\vec{k}) = E_{\vec{k}} - \sum_{\lambda=1}^{2} \alpha_{\lambda} \Omega_{\lambda} + l_{1} \Omega_{1} + l_{2} \Omega_{2} \qquad (l_{1}, l_{2} = 1, 2, \dots, \infty)
\end{equation}
with all possible combinations of numbers of phonon energies, which correspond to the bound states of quasiparticle with superposition of both phonon modes. We should note that if one mode is absent in the system,  one parameter $\alpha_{\lambda}$ becomes equal to zero. Thus, in (\ref{eq14}), in sum over  $\lambda$ there is only one component and sums over $l_{1}$ and $l_{2}$ are equal to zero. Hence, the renormalized spectrum, as it should be, is the same as that obtained for a one-mode system \cite{Dav76,Tka84,Tka16}, and the combined mixed states are absent.

An interesting property of renormalized spectrum (\ref{eq15}) is that if the energy of one mode is  $p$-multiple with respect to the energy of the second mode, then the main and first $p-1$ satellite levels are non-degenerate while the rest are degenerate. Herein, the whole discrete spectrum is equidistant with the same energetic period which is equal to the smallest phonon energy. If the ratio between both phonon energies is a rational number, then the satellite levels which satisfy the condition $p_{1}\Omega_{1}=p_{2}\Omega_{2},\, (p_{1},p_{2}=1,2,\dots,\infty)$ are degenerate while the rest are non-degenerate. If the ratio $\Omega_{1}/\Omega_{2}$ is irrational, then the whole discrete spectrum is non-degenerate.

Renormalized spectrum of the system (\ref{eq15}), the relationship between the Fourier image of retarded Green's function and mass operator $[M(\vec{k},\omega)]$ within Dyson equation \cite{Abr12,Agr68,Dav76}
\begin{equation} \label{eq19}
G(\vec{k}, \omega) = \left\{\omega - E_{\vec{k}} - M(\vec{k},\omega) \right\}^{-1}
\end{equation}
makes it possible to define the average number of phonons $(N_\text{st})$ in the so-called phonon ``dress'' of quasiparticle in stationary  states of the system with energies $E_\text{st}$. It is well known \cite{Lev74,Mis14} that these numbers are fixed by the expression
\begin{equation} \label{eq20}
N_\text{st} = \left\{1 - \left[M'_{\omega}(\omega = E_\text{st}) \right]^{-1} \right\}^{-1} \equiv 1 - G^{2}(\omega = E_\text{st}) \left \{G'_{\omega}(\omega = E_\text{st}) \right \}^{-1}
\end{equation}
and are calculated analytically exactly in this model. As a result, the exact analytical expression is obtained for the average number of the phonons of ground band with $E_{0,0}(\vec{k})$ the energy and two unmixed groups $\lambda=1,2$ of satellite levels $E_{l_{\lambda}}(0)$ of states of quasiparticle bound to each mode, separately
\begin{equation} \label{eq21}
N(\alpha_{1}, \alpha_{2}; l_{\lambda}) = 1 - \alpha^{l_{\lambda}}_{\lambda}(l_{\lambda}!)^{-1} \re^{-(\alpha_{1} + \alpha_{2})}, \, \qquad (l_{\lambda} = 0, 1, 2, \dots, \infty; \, \lambda = 1, 2)
\end{equation}
while for the average number of phonons in mixed groups of satellite states with  $E_{l_{1},l_{2}}(\vec{k})$ energies
\begin{equation} \label{eq22}
N(\alpha_{1}, \alpha_{2}; l_{1}, l_{2}) = 1 - \alpha^{l_{1}}_{1} \alpha^{l_{2}}_{2}(l_{1}! \, l_{2}!)^{-1} \re^{-(\alpha_{1}+\alpha_{2})}, \, \qquad  (l_{1},l_{2}=1,2,\dots,\infty),
\end{equation}
where $l_{1}$ and $l_{2}$ are the numbers of energies ($\Omega_{1}$ and $\Omega_{2}$) of phonon modes which participate  in the formation of the corresponding mixed bound states of the system.

Expressions (\ref{eq21}) and (\ref{eq22}) prove that if there is only one mode in the system (for example, $\Omega_{2}=0, \alpha_{2}=0$), the average number of phonons in mixed groups of the states loses its sense due to their absence. The average number of phonons with the energy $\Omega_{1}$ and coupling constant  $\alpha_{1}$ in ground and satellite states of the system
\begin{equation} \label{eq23}
N(\alpha_{1}; l_{1}) = 1 - \alpha^{l_{1}}_{1}(l_{1}!)^{-1} \re^{-\alpha_{1}}\, \qquad  (l_{1} = 0, 1, 2, \dots, \infty)
\end{equation}
in the limit case is identical to that, obtained for the one-mode system at $T=0$~K \cite{Tka16}.

\begin{figure}[!t]
\centerline{\includegraphics[width=0.65\textwidth]{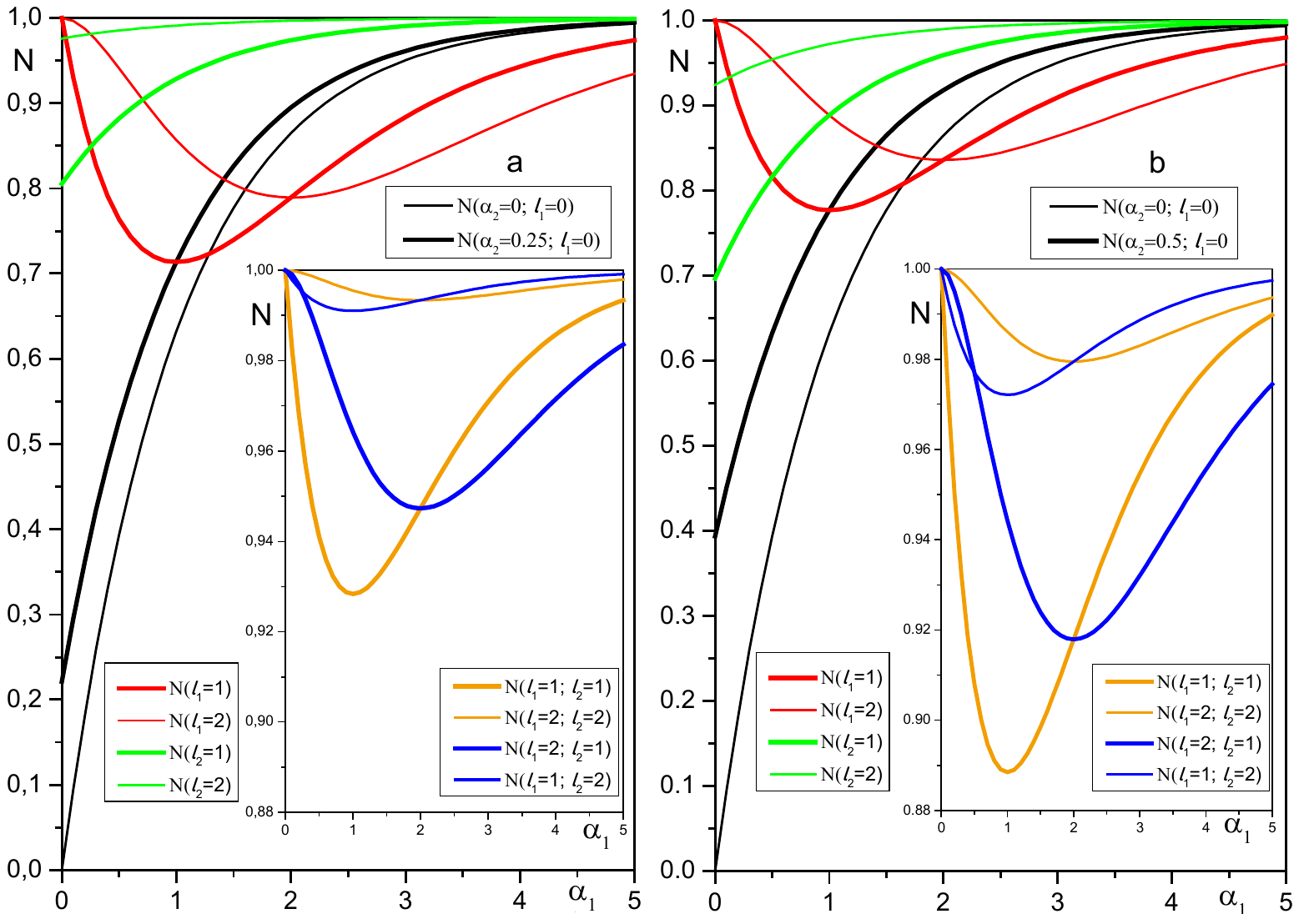}}
\caption{(Colour online) The average number of phonons $N$ as function of  $\alpha_{1}$ at $\alpha_{2}=0.25\, (a)$, $\alpha_{2}=0.5\, (b)$ in the main (at $\alpha_{2}=0\,$ and $\alpha_{2}\neq0)$ and four unmixed phonon satellite states (main panels in the figure) and four mixed states (inserts in the figure).} 
\label{Graf}
\end{figure}

The analysis of these formulae  for arbitrary states and average numbers of phonons $(N)$ calculated as functions of $\alpha_{1}$ at $\alpha_{2} = \textmd{const}$ for several lower renormalized states of the system, presented in figure~\ref{Graf}, show the following. In the ground state with  $E_{0,0}(k=0)$ energy (for example, at $\alpha_{2} =\textmd{const} \geqslant 0$, $l_{1}=l_{2} = 0$), the average number of phonons, as function of $\alpha_{1}$ has analytical form
\begin{equation} \label{eq24}
N(\alpha_{1}, \alpha_{2} = \textmd{const} \geqslant 0, 0) = 1 - \re^{-(\alpha_{1} + \alpha_{2})}.
\end{equation}
Hence, if $\alpha_{1}$ increases from zero to infinity, it monotonously increases from the minimum magnitude $N(0, \alpha_{2}, 0) = 1 - \exp(-\alpha_{2})$ to the maximum one:  $N(\alpha_{1} \rightarrow \infty, \alpha_{2}, 0) \rightarrow 1$. It means that at small values  $(\alpha_{1} + \alpha_{2} \ll 1)$, the average number of phonons in the ``dress'' is small $(N \sim \alpha_{1} + \alpha_{2})$. Thus, the ``dressed'' quasiparticle, in this state, has the properties similar to those, which it had before the interaction with phonons. If the sum of both constants approaches the vicinity where $\alpha_{1} + \alpha_{2} = \ln 2$, the number  $N(\alpha_{1}, \alpha_{2},0)$ is equal to $1/2$. Thus, according to the classification of bound states proposed in \cite{Lev74}, it means that in this state the system has the properties characteristic of the phonon-quasiparticle complex. If the value $(\alpha_{1}+\alpha_{2})$ essentially exceeds $\ln2$, then,  $N(\alpha_{1}, \alpha_{2}, 0) \rightarrow 1$. Therefore, this bound state of the system has predominantly phonon properties.

Functional dependences of the average phonon numbers $N$ on $\alpha_{1}$ at $\alpha_{2} = \textmd{const}$ in both unmixed groups of satellite levels, as it is clear from (\ref{eq21}) and figures~\ref{Graf}a, b, differ from each other due to the different positions of their minima in coordinates $(\alpha_{1}, N)$. The numbers $N(\alpha_{1}, \alpha_{2} = \textmd{const}, l_{1})$ correspond to the group of levels with  $E_{l_{1}}(\vec{k}=0)$ energies. They are minimal at integer $\alpha_{1} = l_{1} = 1, 2, \dots$ values. Thus, their minimum magnitudes are fixed by the expression
\begin{equation} \label{eq25}
\min N(\alpha_{1} = l_{1}, \alpha_{2} = \textmd{const}, l_{1}) = 1 - l_{1}^{l_{1}}(l_{1}!)^{-1} \re^{-(l_{1} + \alpha_{2})}, \qquad (l_{1} = 1, 2, \dots).
\end{equation}

The numbers $N(\alpha_{1}, \alpha_{2} = \textmd{const}, l_{2})$ correspond  to the group of levels with  $E_{l_{2}}(\vec{k}=0)$ energies. All their minima are located at $\alpha_{1}=0$, with their magnitudes given by
\begin{equation} \label{eq26}
\min N(\alpha_{1} = 0, \alpha_{2} = \textmd{const}, l_{2}) = 1 - \alpha^{l_{2}}_{2} (l_{2}!)^{-1} \re^{-\alpha_{2}},\qquad (l_{2} = 1, 2, \dots).
\end{equation}

The properties of the average number of phonons in both unmixed groups of satellite levels is clear from formulae (\ref{eq25}), (\ref{eq26}) and figure~\ref{Graf}a,b. The main property is that in all these bound-to-phonon states of quasiparticle, the system has the properties more similar to the phonons one, the higher is the level in any of both satellite groups.

Finally, the average number of phonons as function of $\alpha_{1}$ at $\alpha_{2}=\textmd{const}$ in all mixed groups of satellite levels with $E_{l_{1},l_{2}}(\alpha_{1},\alpha_{2};l_{1},l_{2})$, energies, as it is clear from (\ref{eq22}) and inserts in figure~\ref{Graf}a, b, is similar to that, corresponding to the group of the levels with $E_{l_{1}}(\vec{k}=0)$ energies, because their minima
\begin{equation} \label{eq27}
\min N(\alpha_{1} = l_{1}, \alpha_{2} = \textmd{const}, l_{1}, l_{2}) = 1 - \frac{l_{1}^{l_{1}} \alpha_{2}^{l_{2}}}{l_{1}! l_{2}!} \re^{-(\alpha_{1}+\alpha_{2})}, \qquad (l_{1} = 1, 2, \dots)
\end{equation}
are also realized at integer  $\alpha_{1}=l_{1}=1,2,\dots$. This fact and figure~\ref{Graf}a, b proves that in all these bound-to-phonon states of quasiparticle, the properties of the system are very similar to the phonon ones.

\section{Main results and conclusions}

The renormalized energy spectrum and the average numbers of phonons are exactly calculated for the model of quasiparticle (in two initial states) strongly interacting with the two-mode phonons at $T=0$~K. Within the exact calculation of Fourier image of retarded Green’s function of quasiparticle, we established that the renormalized spectrum of two-mode system is discrete, as well as that known for the one-mode system \cite{Dav76,Tka84,Tka16}. Furthermore, it does not decay, with respect to physical considerations. The renormalized spectrum of multi-mode (the same for the two-mode) system obtained in \cite{Dav76} at $T=0$~K turned out to  decay due to approximated calculation of the Fourier image of retarded Green’s function. However, from quantum mechanical considerations, interaction of quasiparticle only with virtual phonons should not produce a decay, in order not to violate the law of the energy conservation for the system under study.

It is shown that the renormalized spectrum of the system contains the main band, which is shifted into the low-energy region, and three groups of infinite series of equidistant levels of phonon satellites. Two of them (unmixed) are produced by the bound states of quasiparticle with all numbers  of each phonon mode, separately, while the third one (mixed) is produced by quasiparticle bound with all possible sums of numbers of both phonon modes. The properties of spectrum and the average phonon numbers are analyzed as functions of the coupling constants for the main and all satellite states of the system.

The proposed method of analytical calculation of renormalized spectrum and the average number of phonons can be generalized for the systems with an arbitrary number of modes, which will be done in future.

\newpage

\ukrainianpart

\title{Перенормований спектр квазічастинки, сильно взаємодіючої з двомодовими поляризаційними фононами при $T=0$~K, у моделі з обмеженням на її стани}
\author{Ткач М.В., Сеті Ю.О., Войцехівська О.М., Гутів В.В.}
\address{Чернівецький національний університет імені Юрія Федьковича, вул. Коцюбинського 2, \\58012 Чернівці, Україна}

\makeukrtitle

\begin{abstract}
\tolerance=3000%
Унітарним перетворенням гамільтоніана фреліхівського типу із застосуванням методу функцій Гріна знайдено точний перенормований енергетичний спектр квазічастинки, що сильно взаємодіє з двомодовими поляризаційними фононами при $T=0$~К у моделі системи з обмеженням на число її початкових станів. Також знайдено точні аналітичні вирази середніх чисел фононів у основному і всіх сателітних станах системи та проаналізована їх залежність від величин взаємодії квазічастинки з обома фононними модами.

\keywords квазічастинка, масовий оператор, фонон, спектр

\end{abstract}
\lastpage

\begin{thebibliography}{99}
\bibitem{Abr12} Abrikosov A.A., Gorkov L.P., Dzyaloshinski I.E., Methods of Quantum Field Theory in Statistical Physics, Dover, New York, 2012.
\bibitem{Mah81} Mahan G.D., Many-Particle Physics, Plenum, New York, 1981.
\bibitem{Lev74} Levinson Y.B., Rashba E.I., Soviet Physics Uspekhi, 1974, \textbf{16}, 892--912, \\\doi {10.1070/PU1974v016n06ABEH004097}.
\bibitem{Mis05} Mishchenko A.S., Phys. Usp., 2005, \textbf{48}, 887, \doi{10.1070/PU2005v048n09ABEH002632}.
\bibitem{Mis09} Mishchenko A.S., Phys. Usp., 2009, \textbf{52}, 1193, \doi{10.3367/UFNe.0179.200912b.1259}.
\bibitem{Dev09} Devreese J.T., Alexandrov A.S., Rep. Prog. Phys., 2009, \textbf{72}, 066501,  	 \doi{10.1088/0034-4885/72/6/066501}.
\bibitem{Bon99} Bon\v{c}a J., Trugman S. A.,  Batisti\'{c} I., Phys. Rev. B, 1999, \textbf{60}, 1633, \doi{10.1103/PhysRevB.60.1633}.
\bibitem{Fil05} De Filippis G., Cataudella V., Marigliano Ramaglia V.,  Perron C.A., Phys. Rev. B, 2005, \textbf{72}, 014307,\\ \doi{10.1103/PhysRevB.72.014307}.
\bibitem{Bie16} Bieniasz K., Berciu M., Oles A.M., Acta Phys. Pol. A, 2016, \textbf{130}, 659--663, \doi{10.12693/APhysPolA.130.659}.
\bibitem{Ber06} Berciu M., Phys. Rev. Lett., 2006, \textbf{97}, 036402, \doi{10.1103/PhysRevLett.97.036402}.
\bibitem{Cov07} Covaci L., Berciu M., EPL, 2007, \textbf{80}, 67001, \doi{10.1209/0295-5075/80/67001}.
\bibitem{Kor06} Kornilovitch P.E., Phys. Rev. B, 2006, \textbf{73}, 094305, \doi{10.1103/PhysRevB.73.094305}.
\bibitem{Ebr12}  Ebrahimnejad H.,  Berciu M., Phys. Rev. B, 2012, \textbf{85}, 165117, \doi{10.1103/PhysRevB.85.165117}.
\bibitem{Mol16} M\"{o}ller M. M., Berciu M., Phys. Rev. B, 2016, \textbf{93}, 035130, \doi{10.1103/PhysRevB.93.035130}.
\bibitem{Mar17} Marchand D.J.J., Stamp P.C.E.,  Berciu M., Phys. Rev. B, 2017, \textbf{95}, 035117, \doi{10.1103/PhysRevB.95.035117}.
\bibitem{Pro98} Prokof'ev N.V., Svistunov B.V., Tupitsyn I.S., J. Exp. Theor. Phys., 1998, \textbf{87}, 310, \doi{10.1134/1.558661}.
\bibitem{Mis00} Mishchenko A.S., Prokof'ev N.V., Sakamoto A., Svistunov B.V., Phys. Rev. B, 2000, \textbf{62}, 6317, \\\doi{10.1103/PhysRevB.62.6317}.
\bibitem{Mar10} Marchand D.J.J., de Filippis G., Cataudella V., Berciu M., Nagaosa N., Prokof'ev N.V., Mishchenko A.S., \\Stamp P.C.E., Phys. Rev. Lett., 2010,  \textbf{105}, 266605, \doi{10.1103/PhysRevLett.105.266605}.
\bibitem{Fil12} De Filippis G., Cataudella V., Mishchenko A.S., Nagaosa N., Phys. Rev. B, 2012, \textbf{85}, 094302,\\ \doi{10.1103/PhysRevB.85.094302}.
\bibitem{Mis14} Mishchenko A.S., Nagaosa N.,  Prokof'ev N., Phys. Rev. Lett., 2014, \textbf{113}, 166402, \\\doi{10.1103/PhysRevLett.113.166402}.
\bibitem{Agr68} Agranovich V.M.,  Theory of Exitons, Nauka, Moscow, 1968, (in Russian).
\bibitem{Dav76} Davydov A.S., Theory of Solids, Nauka, Moscow, 1976, (in Russian).
\bibitem{Gio09} Giorgetta F.R., Baumann E., Graf M., Yang Q., Manz C., Kohler K., Beere H.E., Ritchie D.A., Linfield E., Davies A.G., Fedoryshyn Y., Jackel H., Fischer M., Faist J., Hofstetter D., IEEE J. Quantum Electron., 2009, \textbf{45}, 1039, \doi{10.1109/JQE.2009.2017929}.
\bibitem{Str01} Stroscio M.A., Dutta M., Phonons in Nanostructures,  Cambridge University Press, Cambridge, 2001.
\bibitem{Tka84} Tkach N.V., Theor. Math. Phys., 1984, \textbf{61}, 1220, \doi{10.1007/BF01035007}.
\bibitem{Tka16} Tkach M.V., Seti Ju.O., Voitsekhivska O.M., Pytiuk O.Yu., Condens. Matter Phys., 2016, \textbf{19}, 43701,\\ \doi{10.5488/CMP.19.43701}.
\bibitem{Gra84} Gradshtein I.S., Ryzhik I.M., Tables of Integrals, Sums, Series and Products, Nauka, Moscow, 1963, (in Russian).
\end{thebibliography}
\end{document}